\documentclass[prl,aps,twocolumn,amsmath,amssymb,showpacs]{revtex4-1}
\usepackage{graphicx,float,amsmath,amssymb}
\usepackage{latexsym}
\usepackage{mathrsfs}
\usepackage{bm}
\usepackage{bbold}

\usepackage{ulem}

\newcommand{\mean}[1]{\left\langle #1 \right\rangle}
\newcommand{\smean}[1]{\langle #1 \rangle}

\newcommand{\svar}[1]{\langle #1 ^2 \rangle_\mathrm{c}}
\newcommand{\cov}[2]{\left\langle #1  #2 \right\rangle_\mathrm{c}}
\newcommand{\scov}[2]{\langle #1  #2 \rangle_\mathrm{c}}
\newcommand{\ket}[1]{|\kern.3ex#1\kern.3ex\rangle}
\newcommand{\bra}[1]{\langle\kern.3ex #1 \kern.3ex|}

\usepackage[usenames,dvipsnames]{color}

\def\D{\mathrm{d}}                  
 \newcommand{\deriv}[2]{\frac{\mathrm{d}#1}{\mathrm{d}#2}}
 \newcommand{\derivp}[2]{\frac{\partial #1}{\partial #2}}

\renewcommand{\geq}{\geqslant}

\def\tagged{0}  

\newcommand{\eotmt}[1]{(#1)}
\def\D{\mathrm{d}}                  
\def\I{\mathrm{i}}                  

\begin{document}

%%%%%%%%%%%%%%%%%%%%%%%%%%%%%%%%%%%%%%%%%%%%%%%%%%%%%%%%%%%%%%%%%%%%%%%%%%%%%%%%%%%%%%%%%%%%

\title{Dynamics of a tagged monomer: Effects of elastic pinning and harmonic absorption}

\author{Shamik Gupta, Alberto Rosso and Christophe Texier}

\affiliation{Laboratoire de Physique Th\'{e}orique et Mod\`{e}les
Statistiques (UMR CNRS 8626), Universit\'{e} Paris-Sud, Orsay, France}

\begin{abstract}
We study the dynamics of a tagged monomer of a Rouse polymer for
different initial configurations. In the case of free evolution, the
monomer displays subdiffusive behavior with strong memory of the initial
state. In presence of either elastic pinning
or harmonic absorption, we show that the steady state is independent of the initial condition which however strongly
affects the transient regime, resulting in
non-monotonous behavior and power-law relaxation with varying exponents. 
\end{abstract}

\date{\today}


\maketitle

%%%%%%%%%%%%%%%%%%%%%%%%%%%%%%%%%%%%%%%%%%%%%%%%%%%%%%%%%%%%%%%%%%%%%%%%%%%%%%%%%%%%%%%%%

It is known that the dynamics of a mesoscopic particle embedded in a
viscous fluid is Markovian, and well described by the Brownian motion.
The particle mean-squared displacement (MSD) grows diffusively in time as
$2Dt$, where $D$ is the diffusion coefficient. However, in a crowded 
environment of interacting particles, the single particle may display anomalous diffusion. 
Let us consider a long Rouse polymer composed of $L$
monomers connected to their nearest neighbors by harmonic springs of
constants $\Gamma$, and immersed in a good solvent. 
Its global dynamics is Markovian,  
and the center-of-mass diffuses with MSD behaving as $2(D/L)t$.
However, the dynamics of a single tagged monomer is non-Markovian, with
the MSD subdiffusing as $\sqrt{2/(\pi\Gamma)}Db_0\sqrt{t}$ for times
$t \ll L^2/\Gamma$
\cite{deGennes:1971}. Here, $b_0$ encodes the memory of the polymer configuration at $t=0$. 
In particular, if the polymer
at $t=0$ is in equilibrium with the solvent, the dynamics of the tagged monomer
is well described \cite{Krug:1997,Panja:2011,Taloni:2010} by a fractional Brownian motion
(fBm), which generalizes the Brownian motion
to the case of non-independent Gaussian increments 
\cite{Mandelbrot:1968,Kolmogorov:1940}. On the other hand, if the
polymer at $t=0$ is out of equilibrium, the dynamics 
displays {\it aging}, in that the increments are not only correlated (as in fBm), but also drawn from a
Gaussian distribution with a time-dependent variance. 
These non-Markovian processes are relevant for many biological phenomena, such as the unzipping of DNA \cite{Walter:2012}, translocation of
polymers through nanopores \cite{Kantor:2004,Zoia:2009,Panja:2010,Panja:2007}, subdiffusion of
macromolecules inside cells \cite{Szymanski:2009,Weber:2010,Jeon:2012,Allegrini:1998}
and single-file diffusion \cite{Lizana:2010}.

In the above applications, often the tagged particle is subject to
either pinning by an elastic spring or absorption. The first case, e.g., corresponds to employing optical
tweezers to confine specific molecules in order to contrast their dynamical behavior inside the crowded environment of
a cell with that outside \cite{Bertseva:2012}. The second situation
arises when a reactant attached to a single monomer encounters an
external reactive site fixed in space \cite{Guerin:2012,Guerin:2013}.
Moreover, in the problems of polymer translocation and DNA unzipping, the time to translocate
or unzip corresponds to 
the absorbing time of a one-dimensional subdiffusive Gaussian process inside a finite interval with absorbing boundaries.
In general, these problems are investigated numerically either by
molecular dynamics simulations or by
simulation of the underlying Gaussian process
\cite{Dieker,Hartman:2013}. Recently, it has been shown that
subdiffusive Gaussian dynamics can be studied by the fractional Langevin equation
\cite{Jeon:2010,Lizana:2010,Panja:2010-1}. This approach has been
fruitfully used in presence of elastic pinning
\cite{Desposito:2006,Desposito:2009,Grebenkov:2011}, but cannot easily
incorporate absorption. 

In this Letter, we propose a general analytical framework to compute relevant quantities
such as the MSD and the absorbing time distribution of the tagged monomer, for the case of elastic pinning and harmonic absorption.  
These problems are relevant for practical applications:
the pinning by optical tweezers is indeed elastic, while harmonic absorption mimics well a finite interval with absorbing boundaries.
Our approach naturally incorporates the initial condition of the system. 
In the following, we specifically consider a one-dimensional Rouse chain, and mention higher dimensions in the conclusions. 
Our main results, summarized in Table
\ref{table1}, show that while the steady state is independent of the
initial condition, the transient behavior exhibits very strong memory
effects: (i) If a quench in temperature is performed at $t=0$, the MSD
displays a bump in time and converges to the steady state value as a
power law. This behavior, predicted for both pinning and absorption,
could be observed in experiments. (ii) For harmonic
absorption, the absorption time distribution decays exponentially with a 
characteristic time which is independent of the initial condition. Hence, we expect the translocation or the unzipping time to have a distribution with exponential tails, independent of the initial condition of the system. 

\begin{figure}[!ht]
\includegraphics[scale=0.8]{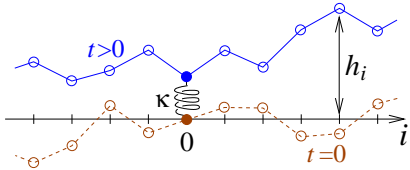}
\caption{(Color online) Schematic of an EW interface pinned by a
harmonic spring acting on the tagged monomer at $i=0$. The initial
configuration $h^0$ (dashed) has $h_0^0=0$.}
\label{fig:polymer-confg}
\end{figure}

The Rouse chain is equivalent to the one-dimensional discrete
Edwards-Wilkinson (EW) interface shown in Fig. \ref{fig:polymer-confg}
\cite{Edwards:1982,note}. Here, $h_i(t)$ is the displacement of the $i$-th monomer at time
$t$ with respect to the origin. The
elastic energy of the system is
$E_\mathrm{el}=(\Gamma/2)\sum_i(h_{i+1}-h_i)^2$, where $\Gamma$ is set to unity below.
Additionally, the monomers are subjected to friction (set to unity) in an overdamped
regime. The dynamics of the interface is described by a set of
$L$ coupled Langevin equations:
\begin{equation}
  \label{eq:eom}
  \frac{\partial h_i(t)}{\partial t}
  =-\frac{\partial E_\mathrm{el}}{\partial h_i} + \eta_i(t)
  =\sum_j \Delta_{ij} h_j(t) + \eta_i(t)
  \:,
\end{equation}
where $\Delta$ denotes the discrete Laplacian matrix, $\{\eta_i(t)\}$ are independent Gaussian white noises: 
$\langle\eta_i(t)\rangle=0$,
$\langle \eta_i(t)\eta_j(t')\rangle=2T\,\delta_{i,j}\,\delta(t-t')$,
with the temperature $T$ set to unity below, and $\langle\cdots\rangle$ denoting thermal averaging.

%%%%%%%%%%%%%%%%%%%%%%%%%%%%%%%%%%%%%%%%%%%%%%%%%%%%%%%%%%%%%

\vspace{0.1cm}
\noindent {\it Elastic pinning.---} 
We consider the situation where the ``tagged" monomer at $i=0$ is pinned around the origin by an additional elastic force (Fig.~\ref{fig:polymer-confg}). This is described by adding the term $\kappa \,h_0^2/2$ to the energy.
In this case, the Langevin equations are similar to \eqref{eq:eom} with
$\Delta_{ij}$ substituted by $-\Lambda_{ij}=\Delta_{ij}-\kappa\,\delta_{i,j}\delta_{i,0}$, 
and 
can be solved (cf. {\it Supplemental Material}). 
In order to adopt a unified formalism to deal with both pinning and absorption, we follow a Fokker-Planck approach.
Let $\mathcal{W}_t[h|h^0]$ be the probability density to observe the
interface in the configuration $h$ at time $t$, given that the configuration at
time $t=0$ was $h^0$, where $h$ (respectively, $h^0$)
denotes the vector $\{h_i\}$ (respectively, $\{h_i^0\}$).  
It obeys the Fokker-Planck equation (FPE)
\begin{equation}
  \label{eq:FFPE}
  \frac{\partial \mathcal{W}_t[h|h^0]}{\partial t}=  
  \Big[
      \sum_i \frac{\partial^2}{\partial h_i^2}
    + \sum_{i,j}\frac{\partial }{\partial h_i} \Lambda_{ij}h_j
  \Big]
  \mathcal{W}_t[h|h^0]
  \:,
\end{equation}
which is a $L$-dimensional generalization of the FPE for the one-dimensional Ornstein-Uhlenbeck process~\cite{Risken:1989,Gardiner:1989}.
Equation \eqref{eq:FFPE} can be solved through a mapping onto the
imaginary time Schr\"odinger equation for $L$ coupled quantum harmonic
oscillators (see {\it Supplemental Material})~:
\begin{eqnarray}
  \label{eq:ConditionalProbaFree2}
  &&\hspace{-0.5cm}\mathcal{W}_t[h|h^0] = 
  \sqrt{ \det\left(\frac{\Lambda}{2\pi(1-e^{-2\Lambda t})}\right) }
  \\ \nonumber 
  &&\hspace{-0.25cm}\times
  \exp\Big[-\frac12(h-e^{-\Lambda t}h^0)^\mathrm{T}
  \frac{\Lambda}{1-e^{-2\Lambda t}}(h-e^{-\Lambda t}h^0)\Big],
\end{eqnarray}
where the superscript "$\mathrm{T}$" denotes transpose operation.
Note that replacing the matrix $\Lambda$ in Eq.~\eqref{eq:ConditionalProbaFree2} by the spring constant $\lambda$, we recover the well-known Ornstein-Uhlenbeck result for the dynamics of a particle submitted to a harmonic force.

Since Eq. \eqref{eq:ConditionalProbaFree2} has a Gaussian form, all
statistical information about the dynamics of the tagged monomer are
encoded in the first two moments of $\mathcal{W}_t[h|h^0]$, which are conveniently obtained by introducing the local field $b=\{b_i\}$ acting on individual monomers. We consider the generating function 
\begin{align}
  \mathcal{G}_t[b]
  = \int \prod_i dh_i\: e^{\sum_i b_i h_i}\,
  \mathcal{W}_t[h|h^0]
  \:.
  \label{eq:GenFunc-nomu}
\end{align}
Using Eq.
\eqref{eq:ConditionalProbaFree2} in Eq. \eqref{eq:GenFunc-nomu},
changing variables $h \to h-e^{-\Lambda t}h^0$, and doing
the Gaussian integration, we get
\begin{align}
  \mathcal{G}_t[b]=\exp\bigg[
   \frac12 b^\mathrm{T} 
  \frac{1-e^{-2\Lambda t}}{\Lambda} b
  + b^\mathrm{T} e^{-\Lambda t} h_0 
  \bigg]
  \:.
\end{align}
\begin{table*}
\begin{tabular}{|c|c|c|c|c|}
 \hline   &Free evolution  & Elastic pinning, & Harmonic absorption,   \\
& & Long time $t \rightarrow \infty$  behavior & Long time $t \rightarrow \infty$  behavior  \\
 \hline  \hline
 Single particle&$\langle h_0^2(t)\rangle = 2 t$ &
 $\langle h_0^2(t)\rangle = \frac{1}{\kappa} (1- \, e^{- 2
 \kappa t})$  &  $\langle h_0^2(t)\rangle =  2 \mu^{-1/2} \, \tanh (t/ \sqrt{\mu}) $     \\
 & Brownian process & &  $ S(t) \approx \exp(- 2 \,  \mu^{1/2} t    ) $ \\
 \hline
 Tagged monomer&  $ \overline{\langle h_0^2(t)\rangle} =
 \sqrt{\frac{2}{\pi}} b_0  \,\sqrt{t}$         &                   $
 \overline{ \langle{h_0^2(t)}\rangle }  =
  \frac{1}{\kappa}  +       \frac{c_0}{ \kappa^2 \sqrt{ t}}
    +\cdots$ ;                            &$ \overline{ \langle
    h_0^2(t)\rangle}^{\rm abs} = a_0 \mu^{-1/3} + O(1/t) $   \\
 $T_0 \ne 1$    & $b_0=1+T_0 (\sqrt{2}-1)$& $c_0 =  \sqrt{2/\pi} (T_0-1) $ &         $   S(t) \approx \exp(-  a_0 \, \mu^{2/3} \, t   ) $    \\
 &   aging process &    &                 \\ \hline
   Tagged monomer&  $\overline{\langle h_0^2(t)\rangle} =
   \frac{2}{\sqrt{\pi}}\sqrt{t}$    &  $\overline{\langle
   h_0^2(t)\rangle} = \frac{1}{\kappa} +  \frac{c_1}{ \kappa^3 t}  +
   \cdots $ ;& $ \overline{\langle h_0^2(t)\rangle}^{\rm abs} =  a_0 \mu^{-1/2}  + O(1/t)$   \\
 $T_0=1$ &    fBm process   &      $c_1 \approx 0.0711 $  &    $ S(t) \approx \exp(-  a_0 \, \mu^{2/3} \, t   ) $                        \\ \hline
\end{tabular}
\caption{Summary of our results for the MSD and the survival probability: single particle versus tagged monomer of an infinite Rouse chain.
We prepare the chain in equilibrium at temperature $T_0$, and the
overbars denote the average over the ensemble of initial configurations. At time $t=0$,
the system is quenched to temperature $T=1$ and let  evolve following
three protocols, namely, (i) free evolution, (ii) elastic pinning
acting on the tagged monomer and (iii) harmonic absorption acting on the
tagged monomer. The friction constant, $\Gamma$ and $D$ are all set to unity. }
\label{table1}
\end{table*}

\noindent
Note that $\mathcal{G}_t[0]=1$ represents the normalization of
$\mathcal{W}_t[h|h^0]$. The connected correlation functions are obtained by differentiation of $\mathcal{F}_t[b]=\ln\mathcal{G}_t[b]$. 
In particular, using $\smean{h_i(t)}=\partial\mathcal{F}_t[b]/\partial b_i\big|_{b=0}$ and 
$\scov{h_i(t)}{h_j(t)}=\partial^2\mathcal{F}_t[b]/\partial b_i\partial b_j\big|_{b=0}$, we get 
\begin{eqnarray}
\hspace{-0.5cm}
\mean{ h_i(t) }= (e^{-\Lambda t}h^0)_i,
\hspace{0.25cm}
\cov{ h_i(t) }{ h_j(t) }=\left(\frac{1-e^{-2\Lambda t}}{\Lambda}\right)_{ij}.  
\label{eq:VarHPinning}
\end{eqnarray}

At long times, we expect from the equipartition theorem that 
$\langle h_0^2(t\to\infty) \rangle = 1/\kappa$, independent of the number of monomers
in the polymer. In the case of a single particle, the steady state value
is reached exponentially fast in time (Table~\ref{table1}). 
For a long polymer, the analysis of Eq.~\eqref{eq:VarHPinning} shows
that the steady state value is reached with a power-law decay where the
exponent depends on the initial configuration.
In particular, we study an initial configuration $h^0$ randomly sampled from
the ensemble of configurations equilibrated at temperature $T_0$ and conditioned on $h^0_0=0$. 
At equilibrium, the displacements $h_i^0$'s are Gaussian distributed as 
$p_{\rm eq}(h^0)=\exp[-\frac12(h^0)^\mathrm{T} \sigma^{-1} h^0]/\sqrt{{\det}(2\pi\sigma)}$, 
where 
$\sigma_{ij}=\overline{h_i^0h_{j}^0}$ is the covariance matrix, 
with overbar denoting averaging with respect to $p_{\rm eq}(h^0)$. 
In the limit $L\to\infty$, the equilibrated EW interface corresponds to two Brownian
trajectories starting at $0$ with diffusion constant equal to $T_0/2$.
The covariance then reads $\sigma_{ij}=T_0\,\theta_\mathrm{H}(ij)\,\mathrm{min}(|i|,|j|)$,
where $\theta_\mathrm{H}(x)$ is the Heaviside function. On the other hand, for a finite interface with periodic
boundary conditions, we have 
$\sigma_{ij}=T_0\Big[\mathrm{min}(i,j)-ij/L\Big]$, where $i,\:j\in\{0,\cdots,L-1\}$.
The computation of $\langle h_0^2 (t)\rangle$ for long times can be performed
analytically in the limit $L \to \infty$. The details are
given in the {\it Supplemental Material}. We get 
\begin{equation}
  \overline{ \smean{h_0^2(t)} }  
  \simeq 
  \frac{1}{\kappa}
  \left[
    1 + \frac{T_0-1}{\kappa}\sqrt{\frac{2}{\pi t}}
    -\frac{T_0c_1}{\kappa^2t}+\cdots
  \right],
  \label{eq:maintext}
\end{equation}
where 
$c_1=0.0711\ldots$.
We thus see that the MSD tends
to the steady state value $1/\kappa$ as
$1/\sqrt{t}$ if $T_0$ is different from unity. 
For $T_0=1$, which corresponds to the temperature of the noise for $t>0$, the
relaxation to steady state is as $1/t$. Moreover, for $T_0>1$, the MSD
has a non-monotonous behaviour in time with a bump. 
This behaviour may be understood as the effect of the large initial
\textit{spatial} fluctuations of the polymer for $T_0>1$ that propagate towards the tagged monomer and increase its \textit{temporal} fluctuations in the transient regime.
Note that the calculation in
Refs.~\cite{Desposito:2006,Desposito:2009,Grebenkov:2011} applies to
polymers equilibrated with the solvent, while here we study the effects of different initial conditions.

%%%%%%%%%%%%%%%%%%%%%%%%%%%%%%%%%%%%%%%%%%%%%%%%%%%%%%%%%%%%%

\vspace{0.1cm}
\noindent{\it Harmonic Absorption.---} 
The FPE is
\begin{equation}
  \label{eq:FPEAbs}
  \hspace{-0.5cm}\derivp{ \mathcal{W}_t[h|h^0] }{ t}=\Big[
     \sum_{i} \derivp{^2}{h_i^2}
     -\sum_{i,j}
     \Big(
        \derivp{}{h_i} \Delta_{ij}h_j 
        + h_i A_{ij} h_j
     \Big)   
     \Big]
   \mathcal{W}_t[h|h^0],
\end{equation}
where the positive definite matrix $A$ describing absorption is
$A_{ij}=\mu\, \delta_{i,j}\, \delta_{i,\tagged}$, with $\mu >0$ being the absorption rate.
Since the absorption probability increases quadratically with distance,
the FPE 
\eqref{eq:FPEAbs} can be solved using
the mapping to a system of coupled quantum harmonic oscillators (details
in {\it Supplemental Material}). We obtain
\begin{align}
  \label{eq:Result1}
  \mathcal{G}_t[b] &= \mathcal{G}_t[0]\,
  \exp\Big[
      b^\mathrm{T}\,\Omega_t^{-1}\,b + b^\mathrm{T}\,\Omega_t^{-1}Y_t\,h^0\Big]
      \:,\\
  \label{eq:Survival}
   \mathcal{G}_t[0] &= \sqrt{\det\left(e^{-t\Delta}Y_t
   \Omega_t^{-1}\right)}
    \exp\Big[-\frac12(h^0)^\mathrm{T} Q_t h^0\Big]
    \:,
\end{align}
where we have introduced the four symmetric matrices 
\begin{align}
  K &= \sqrt{\Delta^2+4A}
  \:, 
  \hspace{0.4cm}
  \Omega_t = K\coth (Kt)-\Delta
  \:,
  \nonumber 
  \\
  Y_t &= {K}/{\sinh (Kt)}
  \:,
  \hspace{0.25cm}
   Q_t=(\Omega_t+2\Delta-Y_t\Omega_t^{-1}Y_t)/2 
  \:. \nonumber 
\end{align}
In presence of absorption, $\mathcal{W}_t[h|h^0]$ is
not normalized to unity, and $\mathcal{G}_t[0]$ is the survival
probability $S(t)$, namely, the probability
that an initial configuration $h^0$ has not been
absorbed upto time $t$ \cite{Redner:2001,Bray:2013}. Note that the
survival probability is the cumulative of the absorbing time distribution. In the long time limit, we have $\Omega_t \approx K-\Delta$ and $Y_t \approx
\exp(-Kt)$, so that the survival probability asymptotically
decays as $S(t) \sim \sqrt{\mathrm{det}(e^{-(K+\Delta)t})}$. Using $\det[\exp(A)]=\exp(\mathrm{Tr}[A])$, we get
\begin{equation}
  \label{eq:SurvivalAsymptotics}
  S(t) \underset{t\to\infty}\sim
  \exp\big[ -t\,\mathrm{Tr}\{K+\Delta\}/2 \big].
\end{equation}
Note that the decay rate is independent of $h^0$. 

Alternatively, one can
obtain an exact expression for $S(t)$ in terms of the tagged monomer MSD, as follows. Using $S(t)=\int \prod_i
dh_i\: \mathcal{W}_t[h|h^0]$, and the FPE 
\eqref{eq:FPEAbs}, we obtain the evolution equation
$\partial_t S(t)=-\mu \, \langle h_0^2(t) \rangle\,
  S(t)$, 
where $\langle \cdots \rangle$ in presence of absorption involves
averaging over surviving realizations only, see Eq.~\eqref{eq:mean-sq-fixedh0} below. 
Using the initial condition $S(0)=1$, the solution is
\begin{equation}
  S(t) = \exp\left( -\mu \int_0^t d \tau\, \langle h_0^2(\tau) \rangle\right). 
\label{eq:St-soln}
\end{equation}

As before, the mean displacement and the connected correlation function are obtained by differentiating the generating function $\mathcal{F}_t[b]=\ln\mathcal{G}_t[b]$; one finds
\begin{align}
\smean{h_i(t)}&=\big(\Omega_t^{-1}Y_t\,h^0\big)_i, ~\scov{h_i(t)}{h_j(t)}=2\,\big(\Omega_t^{-1}\big)_{ij}
\:.
\end{align}
The  correlation function $\scov{h_i(t)}{h_j(t)}$ is independent of the
initial condition $h^0$ and has a finite value in the long time
limit, while $\smean{h_i(t)}$ vanishes in that limit. In particular,
the MSD in the long time
limit reaches a steady state value: $\langle h_0^2(t \to \infty)
\rangle=\left(2/(K-\Delta)\right)_{00}$. 

A dimensional analysis in the limit of a long polymer, $L \to \infty$,
allows to deduce that $\langle h_0^2(t \to
\infty)\rangle=a_0\mu^{-1/3}$, where $a_0$ is a dimensionless constant of order unity. Noting that in absence of absorption, the tagged monomer subdiffuses as $\langle h^2_0(t) \rangle \sim \sqrt{t}$, we see from the absorbing term in the 
FPE 
\eqref{eq:FPEAbs} that absorption is effective over times such that $\mu
\,t^{3/2} \sim O(1)$. Thus, we have 
$\langle h_0^2(t) \rangle \sim \sqrt{t}\,F(\mu t^{3/2})$, 
where the scaling function $F(x)$ is a constant as $x \to 0$.
Since $\langle h_0^2(t \to \infty) \rangle$ approaches a constant, it follows that $F(x \to \infty)\sim
x^{-1/3}$, giving $\langle h_0^2(t \to \infty) \rangle =a_0 \, \mu^{-1/3}$.
Equation \eqref{eq:St-soln} gives $S(t)\sim\exp[-a_0\,\mu^{2/3}t]$ in
the long time limit, independently of $h^0$, see Table~\ref{table1}.

\begin{figure}[ht!]
\includegraphics[width=0.5\textwidth]{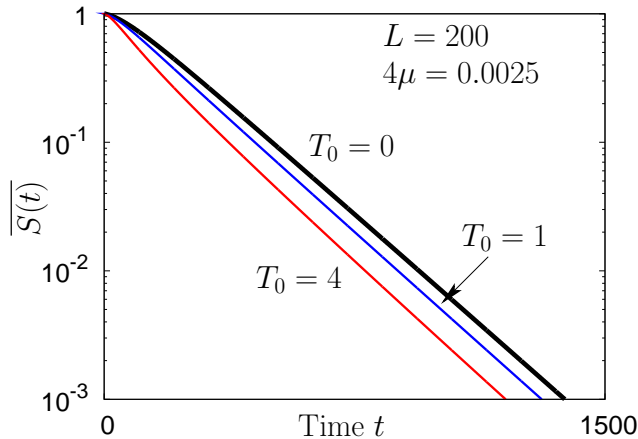}
\caption{(Color online) Survival probability for different initial
temperatures $T_0$. We observe at long times an exponential decay,
$\overline{S(t)}\sim\exp[-a_0\mu^{2/3}t]$, independent of the initial condition. }
\label{Python-results0}
\end{figure}

We now discuss the full time evolution of $\langle h_0^2(t)\rangle$ for
a given initial configuration $h^0$. The MSD is 
\begin{align}
\langle h_0^2(t)\rangle
=\frac{\int \prod_i dh_i\: h_0^2\: \mathcal{W}_t[h|h^0]}{\int \prod_i dh_i\: \mathcal{W}_t[h|h^0]}
\:.
\label{eq:mean-sq-fixedh0}
\end{align}
In order to evaluate the MSD involving an average over an ensemble of
initial configurations, we
should weigh the contribution \eqref{eq:mean-sq-fixedh0} with
$p_{\rm eq}(h^0)S(t)/\overline{S(t)}$, where $S(t)/\overline{S(t)}$ is the probability that the configurations starting from
$h^0$ at time $t=0$ belong to the ensemble of surviving configurations
at time $t$. Denoting the average MSD as 
$\overline{\langle h_0^2(t)\rangle}^{\rm abs}$, we compute it from the generating function 
\begin{align}
  &\ln\big(\, \overline{\mathcal{G}_t[b]}\, \big) 
  = \ln\overline{S(t)}+\frac12 b^\mathrm{T}\, C_t\, b, \\
  \label{eq:MeanSurvival}
  &\overline{S(t)}
  =\sqrt{\frac{\det(e^{-t\Delta}Y_t\Omega_t^{-1}) }{\det(\mathbf{1}+\sigma
  Q_t)}}, \\
  &C_t = 2 \, \Omega_t^{-1} +
  \Omega_t^{-1}Y_t\left(\mathbf{1}+\sigma Q_t\right)^{-1}\sigma\,Y_t\,\Omega_t^{-1},  
\end{align}
with $\mathbf{1}$ the identity matrix.
In particular, we obtain
\begin{equation}
  \label{eq:ResultEqlbmIC1}
  \overline{ \mean{ h_0^2(t) } }^{\rm abs}
  = \partial^2\ln\big(\, \overline{\mathcal{G}_t[b]}\, \big)/\partial b_0^2 \big|_{b=0} 
  = \left( C_t \right)_{00}.
\end{equation}

\begin{figure}[ht!]
\centering
\includegraphics[width=0.5\textwidth]{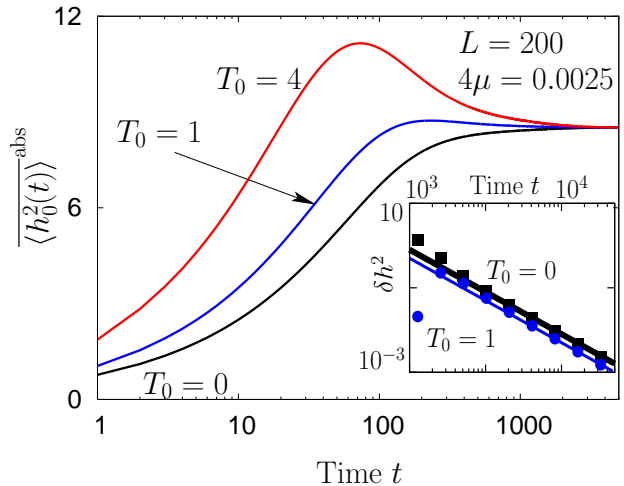}
\caption{(Color online) MSD in presence of harmonic absorption, Eq.~\eqref{eq:ResultEqlbmIC1}. 
The MSD converges to a constant which is independent of $T_0$. 
Inset~: Plot of $\delta h^2=\big|\overline{ \mean{ h_0^2(t) } }^{\rm
abs}-\mean{ h_0^2(\infty)}\big|$ shows the $\sim1/t$ approach to the
steady state.}
\label{Python-results2}
\end{figure}

We compute numerically \eqref{eq:MeanSurvival} and
\eqref{eq:ResultEqlbmIC1} for different initial temperatures $T_0$. 
 The results are shown in Figs.~\ref{Python-results0} and ~\ref{Python-results2}.
As expected by our scaling arguments, both the decay rate of $S(t)$ and
the steady state value of the MSD are independent of the initial
configuration.
For the MSD, the approach to the steady state value $a_0\mu^{-1/3}$ is
always as $1/t$ (inset of Fig.~\ref{Python-results2}), 
i.e. faster than the behaviour $1/\sqrt{t}$ obtained for the case of pinning. 
For initially flat interface (i.e. $T_0=0$), we see from Fig.~\ref{Python-results2} that \eqref{eq:ResultEqlbmIC1} behaves monotonically in time. 
While for elastic pinning, a bump appears only above $T_0=1$,
with absorption 
a bump is observed already for $T_0=1$, and further enhanced for larger $T_0$ (Fig.~\ref{Python-results2}). 
It would be interesting to understand why the approach to steady state differs in the two cases. 
Our numerical results are supported by direct Monte Carlo simulations of the interface dynamics, and by a careful finite-size analysis
presented in the {\it Supplemental Material}.

%%%%%%%%%%%%%%%%%%%%%%%%%%%%%%%%%%%%%%%%%%%%%%%%%%%%%%%%%%%%%

\vspace{0.1cm}
\noindent{\it Conclusion.---} 
In this paper, we analyzed tagged monomer
dynamics under the action of elastic pinning or harmonic absorption. Our solution stems from the crucial observation that in
presence of harmonic interactions, the stochastic evolution of the
tagged monomer remains Gaussian. Some of our results, e.g., the presence
of a unique steady state or the bump in MSD corresponding to a
temperature quench, can be intuitively understood. Others like the
exponential decay of the survival probability or the
power-law transient behaviors in presence of absorption were observed in numerical simulations \cite{Kantor:2004},
but were not analytically known before. Finally, some of our results
like the change of power law for $T_0 \ne 1$ (pinning case) or the
bump observed when $T_0=1$ (harmonic absorption) were
unexpected.

In this work, we focussed on the case of one-dimensional polymers. 
However, it is straightforward to generalize our analysis
to either a Rouse chain in $d$ dimensions \cite{Panja:2013} or a $d$-dimensional EW
interface, by using the corresponding Laplacian matrix in place of
$\Delta$. Moreover, hydrodynamic effects for the chain or long-range
elastic interactions for the interface can also be included by replacing
$\Delta$ with the corresponding fractional Laplacian $-(-\Delta)^{z/2}$ \cite{Krug:1997}; in
this case, the MSD of the tagged particle subdiffuses as $t^{(z-1)/z}$
with $z>1$ for the chain, and as $t^{(z-d)/z}$ with $z>d$ for the
interface \cite{Zoia:2007}, see {\it Supplemental Material}. It would be interesting to study the effect
of the pinning and absorption in the case of non-linear models such as 
self-avoiding polymers, and KPZ interfaces \cite{Gupta:2007}. Another open issue is to go beyond the harmonic approximation and study
absorption in presence of localized targets.

\vspace{0.1cm}
\noindent{\bf Acknowledgements.---}
SG and AR acknowledge CEFIPRA Project 4604-3 for support. 
We thank M. Kardar for very helpful discussions all along this work.

\newpage

\begin{widetext}
\begin{center}
{\bf\large Dynamics of a tagged monomer~: Effects of elastic pinning and harmonic absorption -- \\
Supplemental Material}
\end{center}
\end{widetext}

\setcounter{equation}{18}
\setcounter{figure}{3}

\section{A. Solution of the Fokker-Planck equations}
\label{solution-FFPE}

Here, we provide some details on the solution of the Fokker-Planck
equations~\eotmt{2} and \eotmt{8} of the main text.

\subsection{A.1. Reminder: 1d Ornstein-Uhlenbeck process}

Let us consider the Langevin equation \cite{Risken}
\begin{equation}
  \deriv{x(t)}{t} = -V'(x(t)) + \eta(t),
\end{equation}
describing a particle at position $x(t)$ submitted to a harmonic force $-V'(x)=-\lambda\,x$ and a Langevin force $\eta(t)$ such that $\langle\eta(t)\rangle=0$ and $\langle\eta(t)\eta(t')\rangle=2\,\delta(t-t')$.
The dynamics of the particle may be equivalently described by the Fokker-Planck equation 
\begin{equation}
  \label{app-eq1}
    \derivp{W_t}{t} 
  = \derivp{^2 W_t}{x^2}
  + \derivp{}{x}\left[V'(x)W_t\right]
  \:,
\end{equation}
where $W_t(x|x_0)$ is the (conditional) probability density to find the particle at $x$ at time $t$, given that it was at $x_0$ at initial time.
A convenient way to solve \eqref{app-eq1} is to write
\begin{equation}
  W_t(x|x_0)=Z_t(x|x_0)\sqrt{\frac{P_\mathrm{eq}(x)}{P_\mathrm{eq}(x_0)}},
  \label{app-eq2}
\end{equation}
where
\begin{equation}
  P_\mathrm{eq}(x) = e^{-V(x)} = e^{-\frac12\lambda x^2}
\end{equation}
is the equilibrium distribution (up to a normalization).
The propagator $Z_t(x|x_0)$ satisfies the Schr\"{o}dinger equation in
imaginary time, 
\begin{equation}
  -\derivp{ Z_t(x|x_0)}{ t}=H_0\, Z_t(x|x_0),
  \label{app-eq3}
\end{equation}
where the Hamiltonian 
\begin{align}
  H_0 
  &= -\frac{\partial^2}{\partial x^2} + \frac{1}{4}(V'(x))^2 -\frac{1}{2}V''(x)
  \\
  &= -\frac{\partial^2}{\partial x^2} + \frac{\lambda^2x^2}{4} -\frac\lambda2.
  \label{app-eq4}
\end{align}
describes here a harmonic oscillator.
The shift of energy $-\lambda/2$ makes the ground state energy of $H_0$ zero, which ensures the conservation of the probability in the diffusion problem $\int\D x\,W_t(x|x_0)=1$.
The solution of \eqref{app-eq3} corresponding to the initial condition $Z_0(x|x_0)=\delta(x-x_0)$ is well-known~\cite{Feynmann-Hibbs}~:
\begin{eqnarray}
\label{propagator-sp}
&&Z_t(x|x_0)=\sqrt{\frac{\lambda \, e^{\lambda t}}{4\pi \sinh(\lambda
t)}}
\\ \nonumber 
&&\times\exp\Big[-\frac{\lambda}{4}\Big((x^2+x_0^2)\coth(\lambda t)-\frac{2xx_0}{\sinh(\lambda t)}\Big)\Big]. 
\end{eqnarray}

\subsection{A.2. Multidimensional Ornstein-Uhlenbeck process}

The Fokker-Planck equation \eotmt{2} of the main text can be solved in the
same manner as the one discussed in the preceding section. The
probability density is
\begin{equation}
  \label{eq:NUT}
  \mathcal{W}_t[h|h^0]
  =\mathcal{Z}_t[h|h_0]
  \sqrt\frac{\mathcal{P}_\mathrm{eq}[h]}{\mathcal{P}_\mathrm{eq}[h^0]},
\end{equation}
where the equilibrium distribution now reads
\begin{equation}
  \label{eq:phi0-h}
  \mathcal{P}_\mathrm{eq}[h] = e^{-\frac12 h^\mathrm{T} \Lambda h}.
\end{equation}
For an interface made up of $L$ monomers, the quantum propagator
$\mathcal{Z}_t[h|h_0]$ obeys a Schr\"odinger equation describing $L$ coupled harmonic
oscillators~:
\begin{equation}
  \label{eq:HO}
  \mathcal{H}_0 
=\Big[-\sum_i\frac{\partial^2}{\partial h_i^2}+\frac14 
\sum_{i,j}h_i\,(\Lambda^2)_{ij}\, h_j - \frac12 \mathrm{Tr}\{\Lambda\}\Big].
\end{equation}
The propagator generalizes \eqref{propagator-sp}, and is given by~:
\begin{align}
  &\mathcal{Z}_t[h|h_0] =
    \sqrt{ \det\left( \frac{e^{\Lambda t }\,\Lambda}{4\pi\sinh(\Lambda t)} \right) }
  \nonumber\\
 &\times\exp\Big(-\frac14\Big[
  h^\mathrm{T}\, \Lambda\coth(\Lambda t)\,h 
  + (h^0)^\mathrm{T}\, \Lambda\coth(\Lambda t)\,h^0
  \nonumber\\
  &\hspace{1cm}
  -h^\mathrm{T}\, \frac{\Lambda}{\sinh(\Lambda t)} h^0 
  -(h^0)^\mathrm{T}\, \frac{\Lambda}{\sinh(\Lambda t)} h
  \Big]\Big).
 \label{eq:PropagatorFreeLine}
\end{align}
Substituting the above result into Eq.~\eqref{eq:NUT} leads to
Eq.~\eotmt{3} of the main text.

\subsection{A.3. Harmonic absorption}

The Fokker-Planck equation in the presence of absorption, Eq.~\eotmt{8} of the main text, can be solved using the same procedure as above. Performing the transformation \eqref{eq:NUT}, with $\Lambda=-\Delta$, shows that $\mathcal{Z}_t[h|h_0]$ is now the propagator for the Hamiltonian obtained by adding to \eqref{eq:HO} the term $h^\mathrm{T}Ah$~: 
\begin{equation}
  \mathcal{H} 
=\Big[-\sum_i\frac{\partial^2}{\partial h_i^2}+\frac14 
\sum_{i,j}h_i\,(K^2)_{ij}\, h_j - \frac12 \mathrm{Tr}\{\Delta\}\Big],
\end{equation}
where $K^2=\Delta^2+4A$ (see main text). The propagator may be obtained along the same lines as in the previous subsection. Finally
\begin{align}
  &\mathcal{W}_t[h|h^0] 
  =\sqrt{ \det\left(\frac{e^{-\Delta t}}{4\pi}Y_t\right) }
  \exp\Big(-\frac14 (h^0)^\mathrm{T} Q_t h^0\Big)\nonumber \\
    &\times \exp\Big(-\frac14
       (h-\Omega_t^{-1}Y_t h^0)^\mathrm{T}\,\Omega_t\, (h-\Omega_t^{-1}Y_th^0)
    \Big),
\label{eq:ConditionalProbaAbs}
\end{align}
where the matrices $\Omega_t, Y_t$ and $Q_t$ are defined in the main
text.

%%%%%%%%%%%%%%%%%%%%%%%%%%%%%%%%%%%%%%%%%%%%%%%%%%%%%%%%%%%%%%%%%%%%%%%%%%%%%%%%%%%%%%%%%%
\section{B. Langevin approach}

We point out that \textit{in the absence of absorption}, the dynamics of the line may as well be described within the Langevin approach.
We can write the solution of Eq.~\eotmt{1} of the main text as
\begin{equation}
  h_i(t) = \left( e^{-\Lambda t} \right)_{ij}h_j^0
         + \int_0^t\D\tau\, \left( e^{-\Lambda(t-\tau)} \right)_{ij}\,\eta_j(\tau)
\end{equation}
with implicit summation over repeated indices. Averaging leads to \eqref{eq:MeanH}.
We now consider the covariance matrix
\begin{align}
  \cov{h_i(t)}{h_j(t)}
  = &\int_0^t\D\tau\int_0^t\D\tau'\, \left( e^{-\Lambda(t-\tau)} \right)_{ik}\,
  \nonumber\\
  &\times\mean{\eta_k(\tau)\eta_l(\tau')}\,
  \left( e^{-\Lambda(t-\tau')} \right)_{lj}
\end{align}
Using $\mean{\eta_k(\tau)\eta_l(\tau')}=2\,\delta_{kl}\,\delta(\tau-\tau')$ gives
\begin{align}
  \cov{h_i(t)}{h_j(t)}
  = 2\int_0^t\D\tau\, \left( e^{-2\Lambda(t-\tau)} \right)_{ij}
\end{align}
that leads obviously to Eq.~\eotmt{6} of the main text.

%%%%%%%%%%%%%%%%%%%%%%%%%%%%%%%%%%%%%%%%%%%%%%%%%%%%%%%%%%%%%%%%%%%%%%%%%%%%%%%%%%%%%%%%%%
\section{C. Harmonic pinning: Derivation of Eq.~\eotmt{7} and its generalisation}

The mean-squared displacement of the tagged monomer can be explicitly computed in the continuum limit and when the line has an infinite length.
The interface height $h_x$ then becomes a field of a continuous variable $x$. 
The discrete Laplacian $\Delta$ is replaced by the Laplacian operator, so that
$\Lambda_{x,x'}\to\delta(x-x')\Lambda_x$, with $\Lambda_x=-\Delta_x+\kappa\,\delta(x)$ and $\Delta_x=\D^2/\D x^2$.
The aim of the section is to compute the variance of the tagged monomer displacement [Eq.~\eotmt{7} of the main text] 
\begin{align}
\label{eq:quantity1}
  \langle h_0^2(t)\rangle_c
   = \bra{0} \frac{1-e^{-2\Lambda_x t}}{\Lambda_x} \ket{0}
\end{align}
and analyse the mean displacement 
\begin{align}
  \label{eq:MeanH}
  \langle h_0(t) \rangle = \int\D x\,\bra{0}e^{-\Lambda_x t}\ket{x}\, h_x^0
\end{align}
containing the information about the initial configuration of the line.
In particular, assuming initial configurations sampled from the ensemble equilibrated at temperature $T_0$, leads to consider
\begin{align}
\label{eq:quantity2}
\hspace{-0.25cm}  \overline{ \smean{h_0(t)}^2 } 
  =\int\D x\D x'\,
  \bra{0}e^{-\Lambda_x t}\ket{x} \, \sigma_{x,x'} \, \bra{x'}e^{-\Lambda_x t}\ket{0}
  \:;
\end{align}
the covariance matrix for the infinite line is:
$\sigma_{x,x'}=\overline{h^0_xh^0_{x'}}=T_0\,\theta_\mathrm{H}(xx')\,\mathrm{min}(|x|,|x'|)$,
where $\theta_\mathrm{H}(x)$ is the Heaviside function.
Rewriting the variance as 
\begin{align}
\label{eq:quantity1bis}
  \langle h_0^2(t)\rangle_c
  =\int_0^{2t} \D\tau \bra{0} e^{-\Lambda_x \tau }\ket{0}
\end{align}
shows that all these quantities require to determine the propagator $\bra{x}e^{-\Lambda_x t}\ket{0}$.
Its Laplace transform, the Green's function, is more conveniently analysed~:
\begin{equation}
  G(x,x';\varepsilon) = \bra{x}(\varepsilon+\Lambda_x)^{-1}\ket{x'}
  \:.
\end{equation}
Writing $\Lambda_x=-\Delta_x+V$,
we can obtain its explicit form thanks to the Dyson equation
$
(\varepsilon+\Lambda_x)^{-1}
=(\varepsilon-\Delta_x)^{-1}-(\varepsilon-\Delta_x)^{-1}\,V\,(\varepsilon+\Lambda_x)^{-1}
$,
that takes the simple form
\begin{align}
  \label{eq:Dyson}
   G(x,x';\varepsilon) = G_0(x,x';\varepsilon) - G_0(x,0;\varepsilon)\, \kappa\, G(0,x';\varepsilon)
   \:,
\end{align}
thanks to the local nature of the potential $V$, where
$G_0(x,x';\varepsilon) = \bra{x}(\varepsilon-\Delta_x)^{-1}\ket{x'}$.
Setting $x=0$ in \eqref{eq:Dyson} provides the value of $G(0,x';\varepsilon)$, hence~\cite{Tex11book}
\begin{align}
  \label{eq:UsefulFormula}
    G(x,x';\varepsilon)&= G_0(x,x';\varepsilon) \\\nonumber 
    & - G_0(x,0;\varepsilon)\frac{1}{1/\kappa+G_0(0,0;\varepsilon)}G_0(0,x';\varepsilon).
\end{align}

\subsection{C.1. Normal Laplacian}

Using $G_0(x,x';\varepsilon)=\frac{1}{2\sqrt{\varepsilon}}e^{-\sqrt{\varepsilon}|x-x'|}$ leads to the explicit form
\begin{align}
  \label{eq:GDeltaZ2}
  G(x,x';\varepsilon)                
  &= \frac{1}{2\sqrt{\varepsilon}}\bigg[
    e^{-\sqrt{\varepsilon}|x-x'|}
    -\frac{e^{-\sqrt{\varepsilon}(|x|+|x'|)}}{2\sqrt{\varepsilon}/\kappa+1}
  \bigg].
\end{align}
An inverse Laplace transform yields the propagator:
\begin{equation}
  \bra{x}e^{-\Lambda_x t}\ket{0} = \int_\mathscr{B}\frac{\D\varepsilon}{2\I\pi}\,
  \frac{e^{-\sqrt{\varepsilon}|x|}}{\kappa+2\sqrt{\varepsilon}}
  \, e^{\varepsilon t}
\end{equation}
where $\mathscr{B}$ is the Bromwich contour.
Deforming the contour in order to skirt around the branch cut $\mathbb{R}^-$ gives the useful representation~:
\begin{align}
  \label{eq:PropagatorPinning}
 \bra{x}e^{-\Lambda_x t}\ket{0}
  =\int_0^\infty\frac{\D\varepsilon}{\pi}  
  \frac{2\sqrt{\varepsilon}\cos(\sqrt{\varepsilon}|x|)+\kappa\sin(\sqrt{\varepsilon}|x|)}{\kappa^2+4\varepsilon}e^{-\varepsilon t}
  \:.
\end{align}

We now come back to the computation of Eq.~\eqref{eq:quantity1}.
We first notice that the infinite time result
\begin{equation}
  \mean{h_0^2(\infty)}=\bra{0}\frac{1}{\Lambda_x}\ket{0}
  =G(0,0;0)=\frac{1}{\kappa},
\end{equation}
agrees with the equipartition theorem. The second term of \eqref{eq:quantity1} is obtained
from the propagator \eqref{eq:PropagatorPinning} as 
\begin{align}
  \int_{2t}^\infty\D\tau\,\bra{0}e^{-\Lambda_x\tau}\ket{0}&= \frac{2}{\pi} \int_0^\infty\frac{\D\varepsilon}{\sqrt{\varepsilon}}
  \frac{e^{-2\varepsilon t}}{\kappa^2+4\varepsilon}.
\end{align}
The integral may be related to the complementary error function  
(formula 3.466 of \cite{Gradshteyn:1994}) leading to
\begin{eqnarray}
\label{eq:CorrelatorPinnigFlat}
&&\hspace{-0.5cm}\langle h_0^2(t)\rangle_c
=\frac{1}{\kappa}
\left[
  1- \mathrm{erfc}\left(\frac{\kappa\sqrt{t}}{\sqrt{2}}\right)\,e^{\frac12\kappa^2t}
\right]
\\\nonumber 
&&\hspace{-0.5cm}= \frac{1}{\kappa}
\left[
  1- 
  \sqrt{\frac{2}{\pi}}\frac{1}{\kappa\sqrt{t}}
  \left(
    \sum_{n=0}^N
    (-1)^n\frac{(2n-1)!!}{(\kappa^2t)^n}+\mathcal{R}_N
  \right)
\right],
\end{eqnarray}
where $\mathcal{R}_N$ is the rest of the asymptotic series.
At short time $t\ll\kappa^{-2}$, using
$\mathrm{erfc}(x)\simeq1-2x/\sqrt\pi$ as $x \to 0$, we recover the subdiffusive behaviour $\svar{ h_0(t) } \simeq\sqrt{2t/\pi}$.

We now turn to the computation of \eqref{eq:quantity2}. 
In the long time limit $t\gg1/\kappa^2$, we may neglect the term
$4\varepsilon$ in the denominator of the integrand in Eq.~\eqref{eq:PropagatorPinning}. We find
\begin{equation}
  \label{eq:PropagatorPinningLargeT}
  \bra{x}e^{-\Lambda_x t}\ket{0} \simeq\frac{1}{\sqrt{\pi}\kappa^2t^{3/2}}
  \left[
     1 + \frac{1}{2}\kappa|x|-\frac{x^2}{2t}
  \right]e^{-x^2/(4t)}
  \:.
\end{equation}
Inserted in \eqref{eq:quantity2}, it leads to
\begin{align}
  \label{eq:PinningContribInitialFluct}
  \overline{ \smean{h_0(t)}^2 }
  \simeq 
  T_0\,\left(\frac{1}{\kappa^2}\sqrt{\frac{2}{\pi t}}  -\frac{c_1}{\kappa^3t}\right),
\end{align}
where 
$
  c_1=(8/\pi)\int_0^\infty\D u\,(1-u)
  \big[2u\,e^{-u^2}-\sqrt{\pi}\,\mathrm{erf}(u)\big]e^{-u^2}
\simeq0.0711
$.
Adding \eqref{eq:CorrelatorPinnigFlat} and
\eqref{eq:PinningContribInitialFluct}, we obtain the long time behaviour
of the mean-squared displacement given by Eq.~\eotmt{7} of the main text.

\subsection{C.2. Generalised Edwards-Wilkinson model and fractional Laplacian}

A generalization of the Edwards-Wilkinson model has been proposed in order to study the dynamics of interfaces with a non-standard elastic force \cite{KruKalMajCorBraSir97}. 
In this case, the calculation of \eqref{eq:quantity1} and \eqref{eq:quantity2} involve the fractional Laplacian~\cite{SamKilMar93,Pod99}.
Following the same lines, we must first give the free Green's function that may be written under the form
$G_0(x,x';\varepsilon) = \varepsilon^{1/z-1}\,\Psi(\varepsilon^{1/z}|x-x'|)$ where
\begin{equation}
  \label{eq:PsiDeU}
   \Psi(x) 
   = \int_{-\infty}^{+\infty}\frac{\D q}{2\pi}\frac{e^{\I qx}}{1+|q|^z} 
  \:.
\end{equation}
Its short scale behaviour is $\Psi(x\ll1)\simeq-(1/\pi)\ln x$ for $z=1$ and 
$\Psi(x\ll1)\simeq1/[z\sin(\pi/z)] + a_z\,x^{z-1}$ for $z>1$, with $a_z=1/[2\Gamma(z)\cos({\pi z}/2)]$.
For $z<2$, the function presents a power law decay  
$\Psi(x\gg1)\simeq(1/\pi)\sin(\pi z/2)\Gamma(z+1)\,x^{-z-1}$, whereas it decays 
exponentially for $z\geq2$ as 
$\Psi(x\gg1)\sim\sin[\pi/z+x\cos(\pi/z)]\,\exp[ -x\sin(\pi/z)]$.
Note that \eqref{eq:PsiDeU} may be explicitely computed for even integers. E.g. 
$\Psi(x)= \sin[\pi/4+x/\sqrt2]\,\exp[-x/\sqrt2]/2$
for $z=4$.

Eq.~\eqref{eq:UsefulFormula} gives
\begin{equation}
  G(x,0;\varepsilon) = \frac{\varepsilon^{1/z-1}\,\Psi(\varepsilon^{1/z}|x|)}{1+\Psi(0)\,\kappa\,\varepsilon^{1/z-1}}
\end{equation}
where $\Psi(0)=1/[z\sin(\pi/z)]$.
Setting $x=0$, a Laplace inversion gives the propagator~;
using \eqref{eq:quantity1bis}, we find
\begin{align}
    &\langle h_0^2(t) \rangle_c 
    \label{eq:Integral61}
    = \frac{1}{\pi z}
    \int_0^\infty\D\varepsilon\,
    \frac{1-e^{-2\varepsilon t}}{
       \varepsilon^{2-1/z} - \frac{2\kappa\,\varepsilon }{z\tan(\pi/z)}
       + \frac{\kappa^2\,\varepsilon^{1/z}}{z^2\sin^2(\pi/z)}
    }
    \:.
\end{align}
In the short time limit, or equivalently when the spring constant vanishes, we recover the subdiffusive behaviour as it should,
$\smean{ h_0^2(t) }=\frac{\Gamma(1/z)}{\pi(z-1)}\,(2t)^{1-1/z}$ for $\kappa\to0$.

In the long time limit, the exponential in Eq.~\eqref{eq:Integral61} selects only the term  $\sim\varepsilon^{1/z}$ in the denominator, hence
\begin{align}
  \smean{ h_0^2(t) }_c \underset{t\to\infty}{\simeq}
   \frac1\kappa
   -\frac{z\sin(\pi/z)}{\Gamma(1/z)}\,\frac{1}{\kappa^2(2t)^{1-1/z}}
   +\cdots
\end{align}
We check that this coincides with \eqref{eq:CorrelatorPinnigFlat} for $z=2$.

%%%%%%%%%%%%%%%%%%%%%%%%%%%%%%%%%%%%%%%%%%%%%%%%%%%%%%%%%%%%%%%%%%%%%%%%%%%%%%%%%%%%%%%%%%

\begin{figure}[!ht]
\includegraphics[width=70mm]{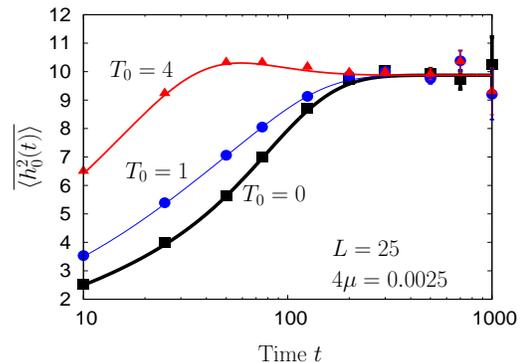}
\caption{(Color online) Monte Carlo simulation (symbols) vs. numerical
evaluation of Eq.~\eotmt{18} of the main text (lines). Simulations involve
average over $10^5$ histories. Each
history starts from an initial configuration drawn from the equilibrium
ensemble at temperature $T_0$. The history contributes to the MSD if
it is not absorbed up to time $t$.}
\label{simulation-results}
\end{figure}

\section{D. Details of Monte-Carlo simulations for the case of harmonic absorption}

Here, we give the details of the Monte Carlo (MC) simulations for the
dynamics of the interface of length $L$ with tagged monomer at $0$
subject to absorption. We start the evolution from the initial $h^0$.
For the equilibrated case, $h_0$ is just a Brownian  bridge implemented  as follows (note that in the program, we have set $h^0_0=h^0_{L-1}=0$ corresponding to $L-1$ independent monomers):
 \begin{align}
 \widetilde{h}_i&= \widetilde{h}_{i-1} + \sqrt{T_0}\, \xi_i, \nonumber \\
 h_i^0&=  \widetilde{h}_i -\frac{i}{L-1}  \widetilde{h}_{L-1}.
\end{align}
Here, $\xi_i$ is a Gaussian distributed random number with zero mean and unit variance, and $T_0$ is the initial temperature.
The covariance matrix is therefore $\sigma_{ij}=T_0\big[\mathrm{min}(i,j)-ij/(L-1)\big]$.

\begin{figure}[!ht]
\includegraphics[width=70mm]{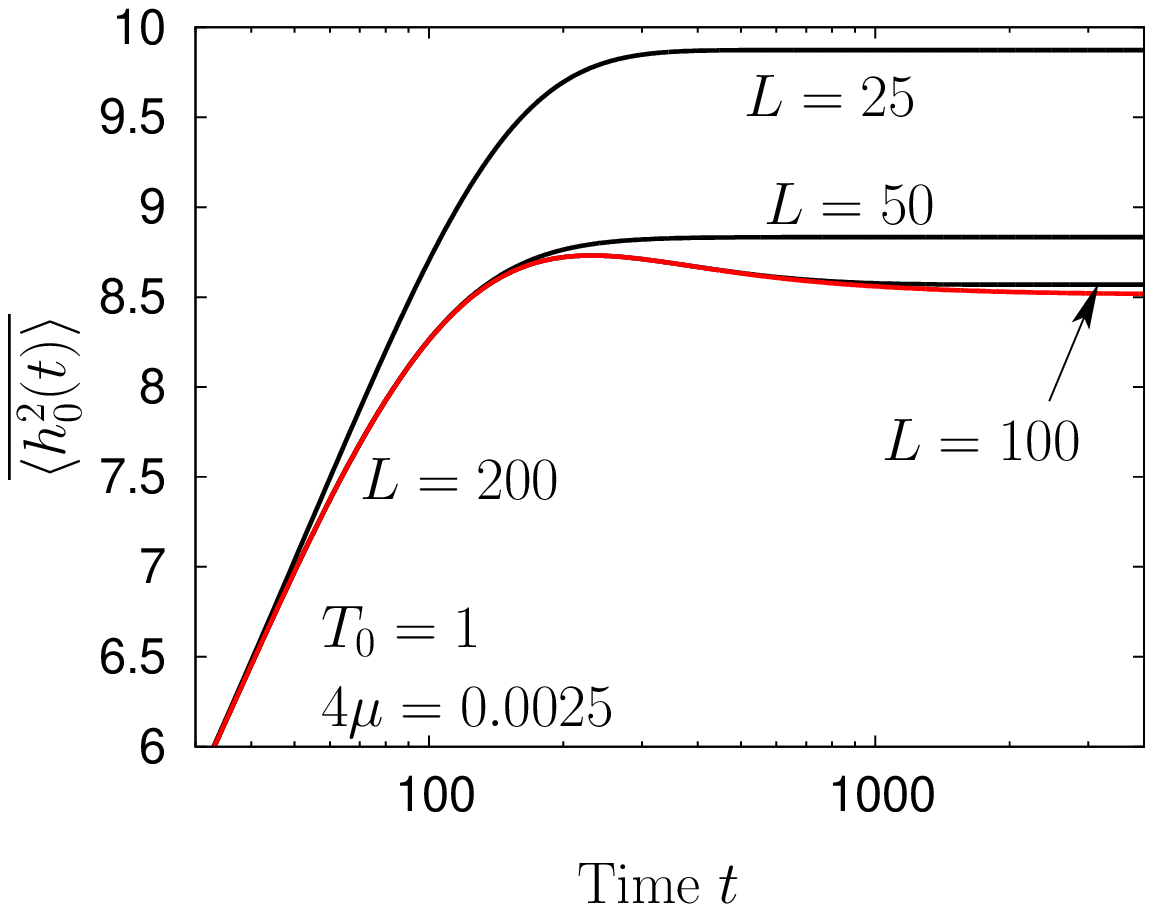}\\
\includegraphics[width=70mm]{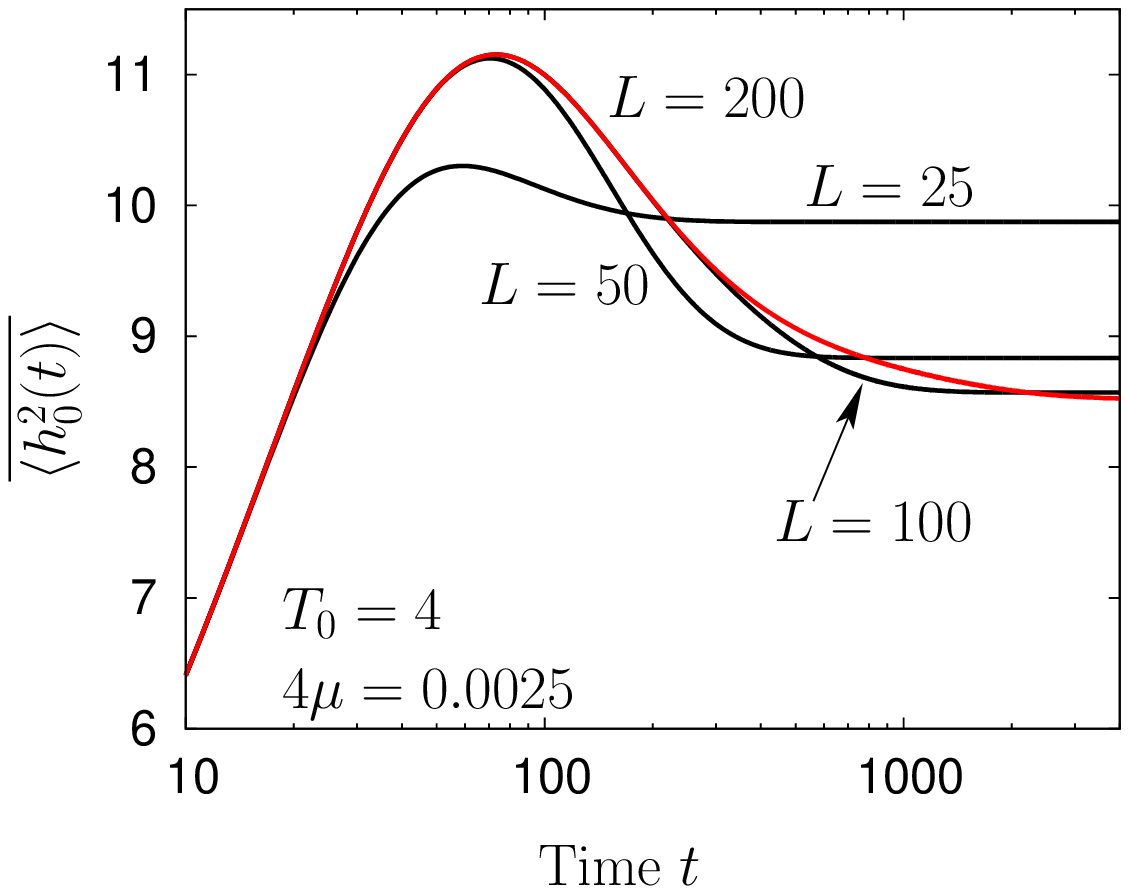}
\caption{(Color online) Finite-size effects at two temperatures,
$T_0=1,4$: Tagged MSD for
different matrix size $L$. The convergence is observed for $L=200$.}
\label{Python-results3}
\end{figure}

Starting from  $h^0$, the
interface configuration is updated between times $t$ and $t+\delta t$
according to:
\begin{eqnarray}
h_i(t+\delta t)&=&h_i(t)+\delta t\left[h_{i+1}(t)+h_{i-1}(t)-2h_i(t) \right] 
  \nonumber\\
  &&+\sqrt{2 \delta t }\, \eta_i(t),
\label{eq:EWupdate}
\end{eqnarray}
for $i=1,2,\ldots,L-1$., while $\delta t \ll 1$ is a
pre-assigned number. Following the update \eqref{eq:EWupdate}, the tagged
monomer  gets absorbed with probability $1-\exp(-\mu h_0^2\delta t)$. In case the tagged monomer is actually absorbed, the whole process of evolving the interface
starts all over again.
Figure \ref{simulation-results} shows MC simulation
results for the variance of the tagged particle displacement, compared with
 numerical evaluation of the matrix defined by Eq.~\eotmt{17} of the main
text; we observe a very good agreement between the two.

Using Eq.~\eotmt{18} of the main text, we can study the limit of long polymers. 
By varying $L$, we show in Fig.~\ref{Python-results3} finite-size effects in the
behavior of the variance of the tagged particle displacement for two
different initial temperatures. In both cases. one observes a convergence in behavior
for $L=200$.


\begin{thebibliography}{100}
\bibitem{note}
In one dimension, the difference between the Rouse chain and the EW
interface is that the displacements are longitudinal in the former and
transversal in the latter. For graphical reasons, we prefer to draw the
interface.
\end{thebibliography}
\end{document}